\documentclass[aps,prb, twocolumn, showpacs,preprintnumbers,amsmath,amssymb,superscriptaddress]{revtex4-1}
        
\usepackage{graphicx}
\usepackage{dcolumn}
\usepackage{bm}
\usepackage{longtable}
\usepackage{color}
\usepackage{textcomp}
\usepackage[utf8]{inputenc} 
\usepackage{hyperref}  
\usepackage{gensymb}
 
\begin{document}     

\preprint{PREPRINT (\today)}

\newcommand{\KFeSe}{K$_x$Fe$_{2-y}$Se$_2$}
\newcommand{\CsFeSe}{Cs$_x$Fe$_{2-y}$Se$_2$}
\newcommand{\RbFeSe}{Rb$_x$Fe$_{2-y}$Se$_2$}
\newcommand{\AFeSe}{$A_x$Fe$_{2-y}$Se$_2$}
 
\title{Magnetic-field tuned anisotropy in superconducting \RbFeSe}    
    
\author{S.~Bosma} 
\email{sbosma@physik.uzh.ch}
\affiliation{Physik-Institut der Universit\"at Z\"urich, Winterthurerstrasse 190, CH-8057 Z\"urich, Switzerland}

\author{R.~Puzniak}
\affiliation{Institute of Physics, Polish Academy of Sciences, Aleja Lotnik\'ow 32/46, PL-02-668 Warsaw, Poland}

\author{A.~Krzton-Maziopa}
\affiliation{Laboratory for Developments and Methods, Paul Scherrer Institute, CH-5232 Villigen PSI, Switzerland}

\author{M.~Bendele} 
\affiliation{Physik-Institut der Universit\"at Z\"urich, Winterthurerstrasse 190, CH-8057 Z\"urich, Switzerland}
\affiliation{Laboratory for Muon Spin Spectroscopy, Paul Scherrer Institute, CH-5232 Villigen PSI, Switzerland}

\author{E.~Pomjakushina}
\affiliation{Laboratory for Developments and Methods, Paul Scherrer Institute, CH-5232 Villigen PSI, Switzerland}

\author{K.~Conder}   
\affiliation{Laboratory for Developments and Methods, Paul Scherrer Institute, CH-5232 Villigen PSI, Switzerland}

\author{H.~Keller}
\affiliation{Physik-Institut der Universit\"at Z\"urich, Winterthurerstrasse 190, CH-8057 Z\"urich, Switzerland}
 
\author{S.~Weyeneth}
\email{wstephen@physik.uzh.ch}
\affiliation{Physik-Institut der Universit\"at Z\"urich, Winterthurerstrasse 190, CH-8057 Z\"urich, Switzerland}
 
\begin{abstract}

The anisotropic superconducting properties of a \RbFeSe\ single crystal with $T_{\rm c}\simeq32$~K were investigated by means of SQUID and torque magnetometry, probing the effective magnetic penetration depth $\lambda_{\rm eff}$ and the penetration depth anisotropy $\gamma_\lambda$. Interestingly, $\gamma_\lambda$ is found to be temperature independent in the superconducting state, but strongly field dependent: $\gamma_\lambda(0.2~\rm T)<4$ and $\gamma_\lambda( 1.4~\rm T)>8$. This unusual anisotropic behavior, together with a large zero-temperature $\lambda_{\rm eff}(0) \simeq 1.8~\mu$m, is possibly related to a superconducting state heavily biased by the coexisting antiferromagnetic phase. 

\end{abstract}
  
 
\maketitle

\section{Introduction}

With the discovery of superconductivity in LaFeAsO$_{1-x}$F$_y$,\cite{Kamihara2008} a new family of iron-based high-temperature superconductors was found. Its simplest member is FeSe$_{1-x}$, which consists of a stack of FeSe layers.\cite{Hsu2008} Its superconducting transition temperature $T_{\rm c}\simeq 8$~K increases drastically with external pressure, reaching $T_{\rm c}(8~{\rm GPa})\simeq 36$~K.\cite{Hsu2008, Margadonna2009} Interestingly, a similar high $T_{\rm c}\simeq 30~$K is attained in the iron-selenide family \AFeSe\ by intercalating alkali atoms ($A$ = K, Rb, Cs) between the FeSe layers.\cite{Guo2010, Krzton-Maziopa2011, Li2011} However, $T_{\rm c}$ is found to decrease with pressure and is fully suppressed at 9~GPa for \KFeSe\ (Ref.~\onlinecite{Guo2011}) and at 8~GPa for \CsFeSe.\cite{Seyfarth2011} The critical temperature is almost insensitive to pressure below 1~GPa, suggesting that $T_c$ is almost independent of small variations of the lattice constants. This provides an opportunity to study the temperature dependence of physical quantities without being affected by changes of the lattice constants due to thermal expansion. Early $\mu$SR experiments on \CsFeSe\ indicate that superconductivity and antiferromagnetism coexist on microscopic length scales.\cite{Shermadini2011} The N\'eel temperature $T_{\rm N}\approx 500$~K (Ref.~\onlinecite{Liu2011}) of K$_{0.8}$Fe$_{2-y}$Se$_2$ is substantially higher than $T_{\rm c}\simeq 30$~K. Several experiments point towards nanoscale phase separation between superconducting, vacancy-disordered domains and vacancy ordered antiferromagnetic (AFM) domains.\cite{Ricci2011, Shen2011, Ksenofontov2011, Wang2011a, Shen2011} In contrast to the slightly hole doped FeSe$_{1-x}$, the intercalation of alkali ions $A$ into the FeSe structure introduces a large amount of electrons into the system.\cite{Ivanovskii2011} Other experiments\cite{Zhang2011} suggested that this highly electron doped system contains no hole-like sheets at the Fermi surface, and thus electron scattering between hole and electron-like bands is impossible.\cite{Ivanovskii2011} Moreover, Fe vacancies in the crystalline structure order at $\simeq 600$~K.\cite{Bao2011} In K$_x$Fe$_{2-y}$Se$_2$, the presence of vacancies appears detrimental to superconductivity.\cite{Li2011a} This intriguing microscopic coexistence of vacancy ordering, antiferromagnetism, and superconductivity in \AFeSe\ points to an unconventional thermodynamic behavior of the superconducting state. \\
\indent Recently, the lower critical field $H_{\rm c1}$ was investigated in tetragonal \KFeSe\ for magnetic fields $H$ along different crystallographic directions, {\it i.e.}~parallel to the $ab$-plane and parallel to the $c$-axis,\cite{Lei2011, Tsindlekht2011} revealing a surprisingly low and isotropic $\mu_0H_{\rm c1}\simeq0.3$~mT. Invoking the phenomenological relation between $H_{\rm c1}$ and the effective magnetic penetration depth $\lambda_{\rm eff}$ for an isotropic superconductor $\mu_0H_{\rm c1}=\Phi_0(\ln{\kappa} + 0.5)/(4\pi\lambda_{\rm eff}^2)$,\cite{Klemm1980} a value of $\lambda_{\rm eff}(0)\simeq1.6-1.8$~$\mu$m is obtained (assuming an approximate Ginzburg-Landau parameter $\kappa\sim100-200$). This low field estimate of $\lambda_{\rm eff}$ deviates remarkably from the small in-plane magnetic penetration depth $\lambda_{ab}(0)\simeq0.29$~$\mu$m derived from NMR experiments at 8.3~T on a similar sample.\cite{Torchetti2011} Moreover, the numerous observations of an anisotropic vortex lattice in high magnetic fields\cite{Torchetti2011, Mun2011, Lei2011, Tsurkan2011} contrast with the isotropic behavior of $H_{\rm c1}$ in low magnetic fields. In order to illuminate this intriguing field dependence of the superconducting properties, we performed a detailed magnetic study of \RbFeSe.
\begin{figure}[t!]
\includegraphics[width = 1\linewidth]{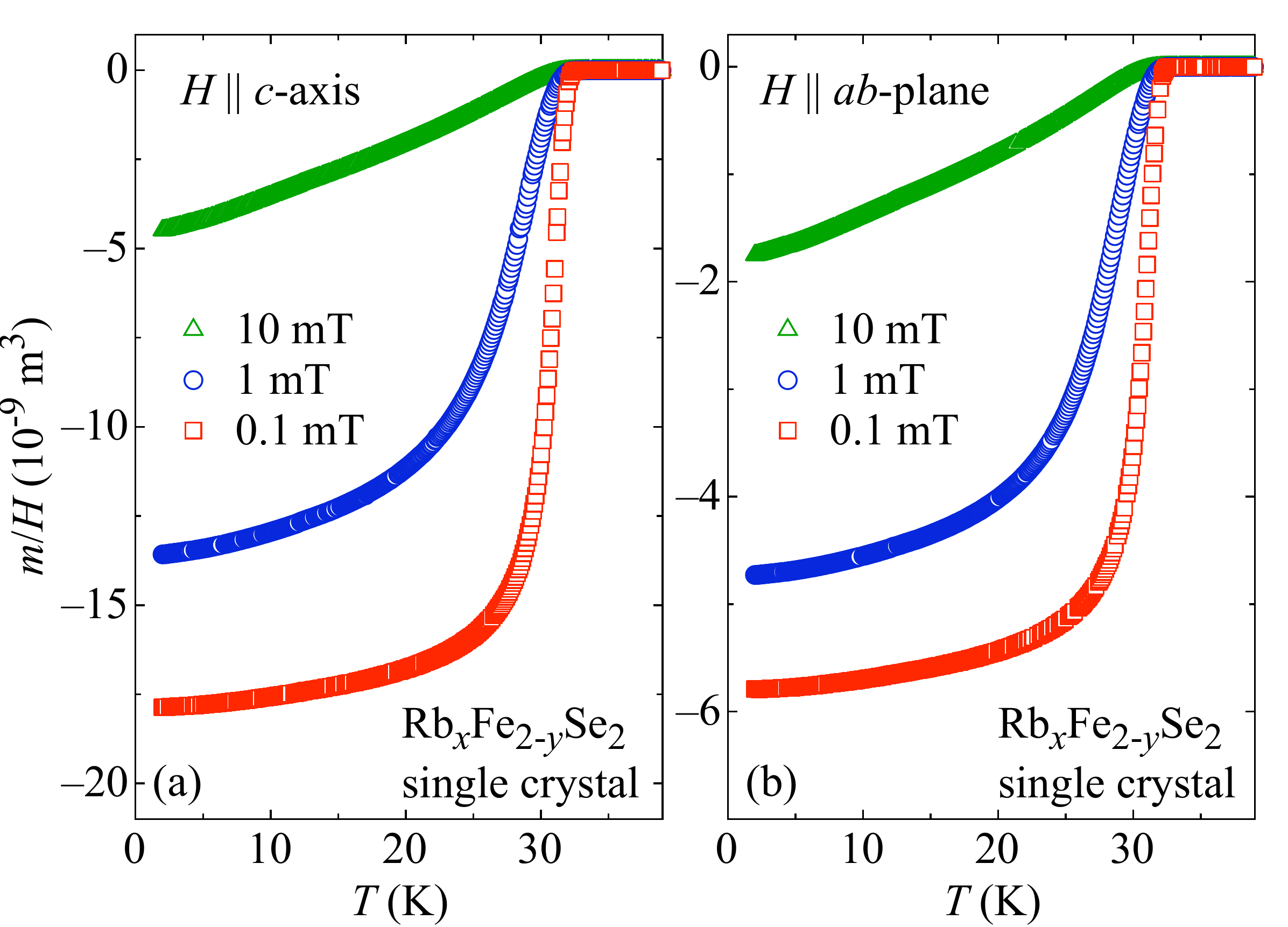}
\caption{(Color online) Zero-field cooled measurements of $m/H$ of single crystal \RbFeSe\  for  (a) $H||c$-axis and (b) $H||ab$-plane. The change of vortex penetration with temperature is very similar for both directions, consistent with an isotropic $\mu_0H_{\rm c1}\lesssim 0.3$~mT. The sharp transition at $T_{\rm c}\simeq32$~K demonstrates the high quality of the crystal. Due to demagnetization the magnitude of $m/H$ varies by a factor of $\sim3$ for both orientations. A rough estimation of the demagnetization factors from the sample dimensions yields $ 0.05<N^{|| ab}<0.3$ and $N^{||c} \sim 0.65$.}
\label{m_TH}
\end{figure} 

\section{Crystal growth}

\indent A \RbFeSe\ single crystal with composition Rb$_{0.77(2)}$Fe$_{1.61(3)}$Se$_2$ as refined by X-ray fluorescence was grown from a melt by the Bridgman method, using a presynthesised ceramic precursor of FeSe$_{0.98}$ and metallic rubidium. For the precursor synthesis high purity (at least 99.99\% Alfa) powders of iron and selenium were mixed in the molar proportion 1 Fe : 0.98 Se and pressed into a rod. This nominal stoichiometry, chosen on the basis of our previous studies\cite{Pomjakushina2009} of the Fe-Se chemical phase diagram, provides an iron selenide of pure tetragonal phase. The mixture was prereacted in a sealed silica ampoule at 700\,°C for 15 hours and then grounded in an inert atmosphere, pressed again into a rod, sealed in an evacuated double wall quartz ampoule and resintered at 700\,°C. After 48 hours the furnace was cooled down to 400\,°C and kept at this temperature for 36 hours more. For the single crystal growth a piece of the Fe-Se rod was sealed in an evacuated silica Bridgman ampoule with an appropriate amount of pure alkali metal placed in an additional thin silica tube; 5\% excess of Rb was added to compensate its loss during synthesis. The Bridgman ampoule was sealed in another protecting evacuated quartz tube. The ampoule was heated at 1030\,°C for 2 hours for homogenization, followed by cooling down the melt to 750\,°C at a rate of 6\,°C/h. Finally, the furnace was cooled down to room temperature at a rate of 200\,°C/h. After synthesis the ampoules were transferred to a He-glove box and opened there to protect the crystal from oxidation in the air. 

\section{Magnetic measurements}

\begin{figure}     
\includegraphics[width = 1\linewidth]{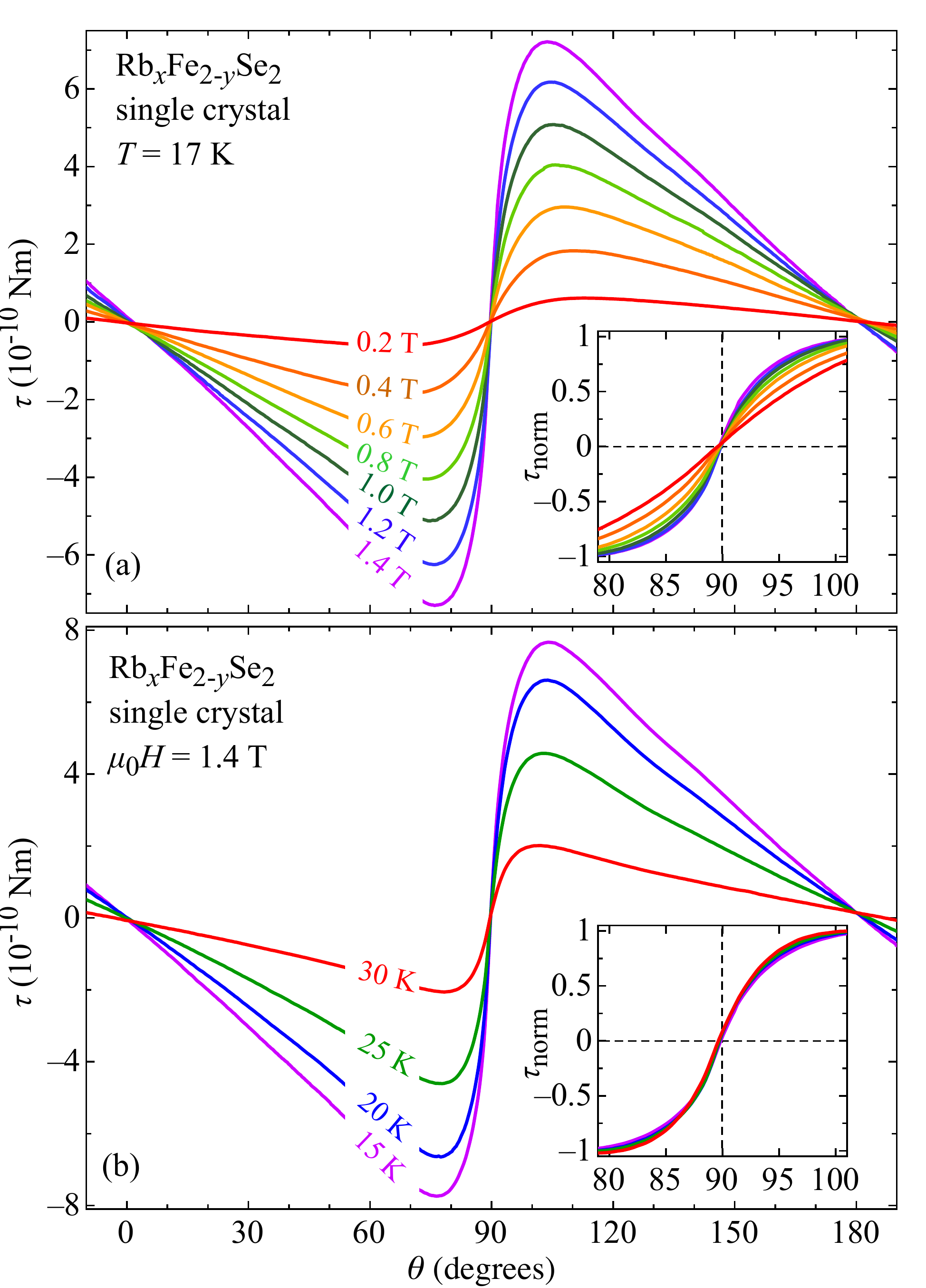}
\caption{(Color online) Angular dependence of the superconducting component of the magnetic torque of \RbFeSe. (a) Magnetic torque at 17~K for various magnetic fields. (b) Magnetic torque at 1.4~T for various temperatures. The insets in both panels show the evolution of $\tau_{\rm norm}$ close to the $ab$-plane.}
\label{torqueTH}
\end{figure}

\indent The superconducting properties of the plate-like crystal of dimensions $\sim5\times1\times 0.2$~mm$^3$ (thickness of 0.2 mm along the $c$-axis) were characterized with a {\it Quantum Design} MPMS XL SQUID magnetometer. The temperature dependence of $m$/$H$, where $m$ is the magnetic moment is shown in Fig.~\ref{m_TH}  for various magnetic fields $H$ applied after zero-field cooling. The onset transition temperature for this sample is estimated to be $T_{\rm c}\simeq 32$~K. For both studied orientations (parallel to the $c$-axis shown in Fig.~\ref{m_TH}a, and parallel to the $ab$-plane shown in Fig.~\ref{m_TH}b), the magnetic properties are very similar. For $\mu_0H=0.1$~mT a sharp diamagnetic transition is observed for both field orientations. In higher fields, the diamagnetism is rapidly suppressed, indicating that $\mu_0H_{\rm c1}$ for \RbFeSe\ is very low ($\mu_0H_{\rm c1} \lesssim$ 0.3~mT, including demagnetizing field correction.\cite{Osborn1945}) \\
\begin{figure}[b!]
\includegraphics[width = 1\linewidth]{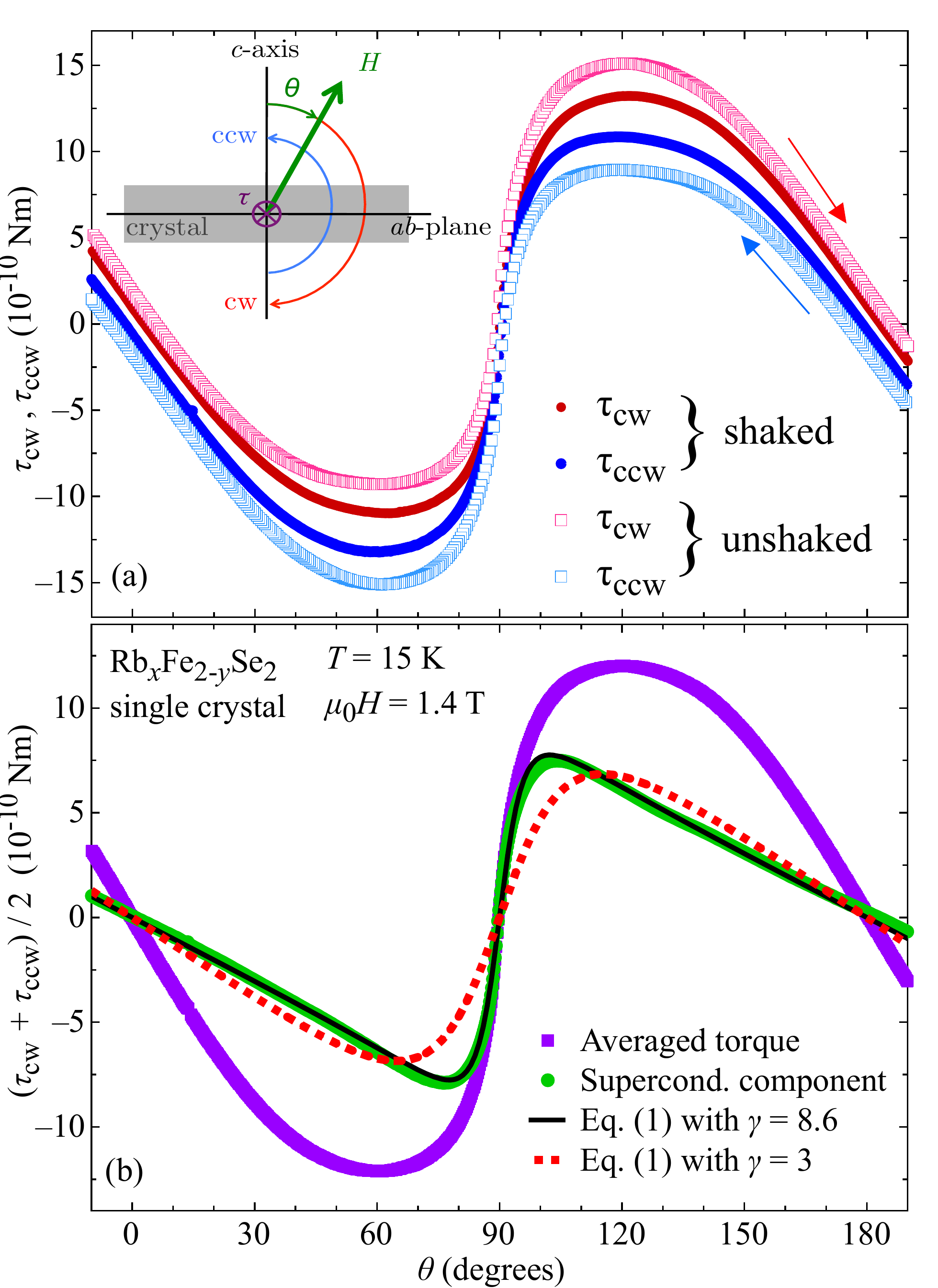}
\caption{(Color online) Angular dependent torque of single-crystal \RbFeSe\ measured at 15 K and 1.4 T. (a) Comparison of unshaked and shaked magnetic torque. The unshaked red data $\tau_{\rm cw}$ (blue data $\tau_{\rm ccw}$) are gathered by turning $H$ clockwise (counter-clockwise) around the sample. The inset explains schematically the field configuration during the experiment. The shaked data are obtained by applying a transverse AC field, yielding reduced irreversibility and enhanced data quality. (b) Angular dependence of the shaked magnetic torque averaged for both directions. The green curve is the superconducting component of the signal, obtained after subtracting the background as explained in the text. A fit by Eq.~(\ref{kogan}) yields $\gamma_\lambda(15~{\rm K},1.4~{\rm T})\simeq8.6$ (black line). For comparison, the red dotted line is calculated with a fixed $\gamma_\lambda = 3$.}
\label{raw}
\end{figure} 
\indent A small piece of of rectangular shape and approximate dimensions $150 \times 150 \times 90~\mu$m$^3$ was cleaved off the above \RbFeSe\ single crystal. Magnetic torque investigations were carried out using a home-made magnetic torque sensor.\cite{Kohout2007} For a uniaxial superconductor like tetragonal \RbFeSe, the angular dependence of the magnetic torque $\vec{\tau}=\vec{m} \times \mu_0 \vec{H}$  in the London approximation ($H_{c1} < H < H_{c2}$) can be written according to\cite{Kogan1988}
\begin{equation}\label{kogan}
\tau(\theta)=-\frac{V\Phi_0H\gamma_\lambda^{2/3}}{16\pi\lambda_{\rm eff}^2}\left(1-\frac{1}{\gamma_\lambda^2}\right)\cdot\frac{\sin(2\theta)}{\epsilon(\theta)}\ln\left(\frac{\eta H_{\rm c2}^{||c}}{\epsilon(\theta)H}\right).
\end{equation}
Here $\theta$ is the angle between $H$ and the crystallographic $c$-axis, $\epsilon(\theta)=[\cos^2(\theta)+\gamma_\lambda^{-2}\sin^2(\theta)]^{1/2}$ is the angular scaling function, $H_{\rm c2}^{||c}$ is the $c$-axis upper critical field, and $\eta$ is a dimensionless parameter of the order of unity. Without loss of generality $\eta$ is restricted to 1 within this work.\cite{Bosma2011} The anisotropy parameter $\gamma_\lambda=\lambda_c$/$\lambda_{ab}$ is the ratio of the out-of-plane and in-plane magnetic penetration depths, whereas $\lambda_{\rm eff}=(\lambda_{ab}^2\lambda_c)^{1/3}$ denotes the effective magnetic penetration depth. It is possible that our sample presents phase separation between superconducting non-magnetic regions and antiferromagnetic, non-superconducting regions, with a domain size smaller than the penetration depth.\cite{Ricci2011} In that case, applying the Kogan model yields a $\lambda_{\rm eff}$ which may be renormalized to a larger value than the superconducting parameter $(\lambda_{ab}^2\lambda_c)^{1/3}$. The field penetrates more easily in this phase separated material, and the Kogan model yields an ``averaged'' effective bulk penetration depth. The good agreement between Eq.~(\ref{kogan}) and the data (Fig.~\ref{raw}b) shows that the Kogan model is still useful, albeit with a broader interpretation of its parameters. \\
\indent The magnetic torque experiments at various $T$ and $H$ were performed by turning $H$ around the sample in a plane containing the $c$-axis (see inset of Fig.~\ref{raw}a), and measuring the resulting torque. As an example, Fig.~\ref{raw}a shows the angular dependence of the torque signal measured at 15 K in 1.4 T. Note that the torque signal is affected by an angular irreversibility between the clockwise ($\tau_{\rm cw}$) and counter-clockwise ($\tau_{\rm ccw}$) branches. Such irreversible angular dependent torque signals are usually observed in hard superconductors due to vortex pinning.\cite{Weyeneth2009a, Weyeneth2009b} Due to the tetragonal structure of \RbFeSe, twinning in the crystal manifests itself in the results only through pinning on the twin boundaries, and all in-plane parameters are not separated along $a$- and $b$- axes in the analysis. In this work, the so-called vortex-shaking technique\cite{Willemin1998} was successfully applied to reduce irreversibility, allowing a more reliable determination of the superconducting parameters. This was done by applying a small AC field orthogonal to $H$ in order to enhance the vortex relaxation towards thermodynamic equilibrium. As seen in Fig.~\ref{raw}a the vortex shaking clearly reduces the irreversible component, especially for $H$ close to the $ab$-plane. In Fig.~\ref{raw}b the average torque $\tau=(\tau_{\rm cw}+\tau_{\rm ccw})/2$ is presented.  The torque signal consists of a superconducting component described by Eq.~(\ref{kogan}) and a magnetic background component $\tau_{\rm BG}=-(\chi_{ab}-\chi_{c})V\mu_0H^2\sin(2\theta)/2$. The variables $\chi_c$ and $\chi_{ab}$ denote the magnetic susceptibilities along the crystallographic axes. Here $\chi_{ab}>\chi_{c}$ (as also mentioned in Ref.~\onlinecite{Torchetti2011}) and $\tau_{\rm BG}$ is large, consistent with a bulk antiferromagnetic phase having the magnetic moments aligned along the $c$-axis.\cite{Shermadini2011, Bao2011} The subtraction of this antiferromagnetic background can be either performed by adding a sinusoidal component in the fitting routine of Eq.~(\ref{kogan}), or by directly removing the symmetric sinusoidal component of the data as discussed in Ref.~\onlinecite{Weyeneth2009b}. All results presented in this work are independent of the background treatment.\\
\indent The parameters $H^{|| c}_{\rm c2}$, $\gamma_\lambda$, and $\lambda_{\rm eff}$ can be extracted simultaneously by analyzing the magnetic torque data with Eq.~(\ref{kogan}). However, in order to reduce the amount of free fit parameters, $H_{\rm c2}^{||c}$ was fixed according to a Werthamer-Helfand-Hohenberg (WHH) temperature dependence.\cite{Werthamer1966} The slope $dH^{|| c}_{\rm c2}/dT$ at $T_{\rm c}$ was fixed to $\sim-1$~T/K, in concordance with resistivity results of a similar sample.\cite{Tsurkan2011} Small variations in $dH^{|| c}_{\rm c2}/dT$ do not affect the results of the analysis, since $H_{\rm c2}$ contributes only logarithmically in Eq.~(\ref{kogan}) and has no weight in the determination of $\gamma_\lambda$ and $\lambda_{\rm eff}$ in low magnetic fields.\cite{Weyeneth2009b} Magnetic torque curves with the antiferromagnetic background subtracted are presented in Fig.~\ref{torqueTH} for various $T$ (Fig.~\ref{torqueTH}a) and $H$ (Fig.~\ref{torqueTH}b). The insets show the normalized magnetic torque $\tau_{\rm norm}=\tau(\theta)/\max[\tau(\theta)]$ close to the $ab$-plane. A change of the shape of $\tau_{\rm norm}(\theta)$ qualitatively reflects a change of $\gamma_\lambda$. Note that the slope of $\tau_{\rm norm}$ vs. $\theta$ changes strongly with $H$, but not with $T$. This demonstrates that $\gamma_\lambda$ is field dependent, but not temperature dependent. As the field direction is approaching the $ab$-plane, the screening currents start to flow not only in plane, but also out of plane, which makes the torque depend strongly on $\lambda_{ ab}$ {\it and} $\lambda_{ c}$ in this angular region. The temperature and field effect on the anisotropy $\gamma_\lambda$ are consequently already visible on the data taken around the $ab$-plane, independently of the strict validity of Eq.~(\ref{kogan}) in a phase separated material. \\
\begin{figure}  
\includegraphics[width = 1\linewidth]{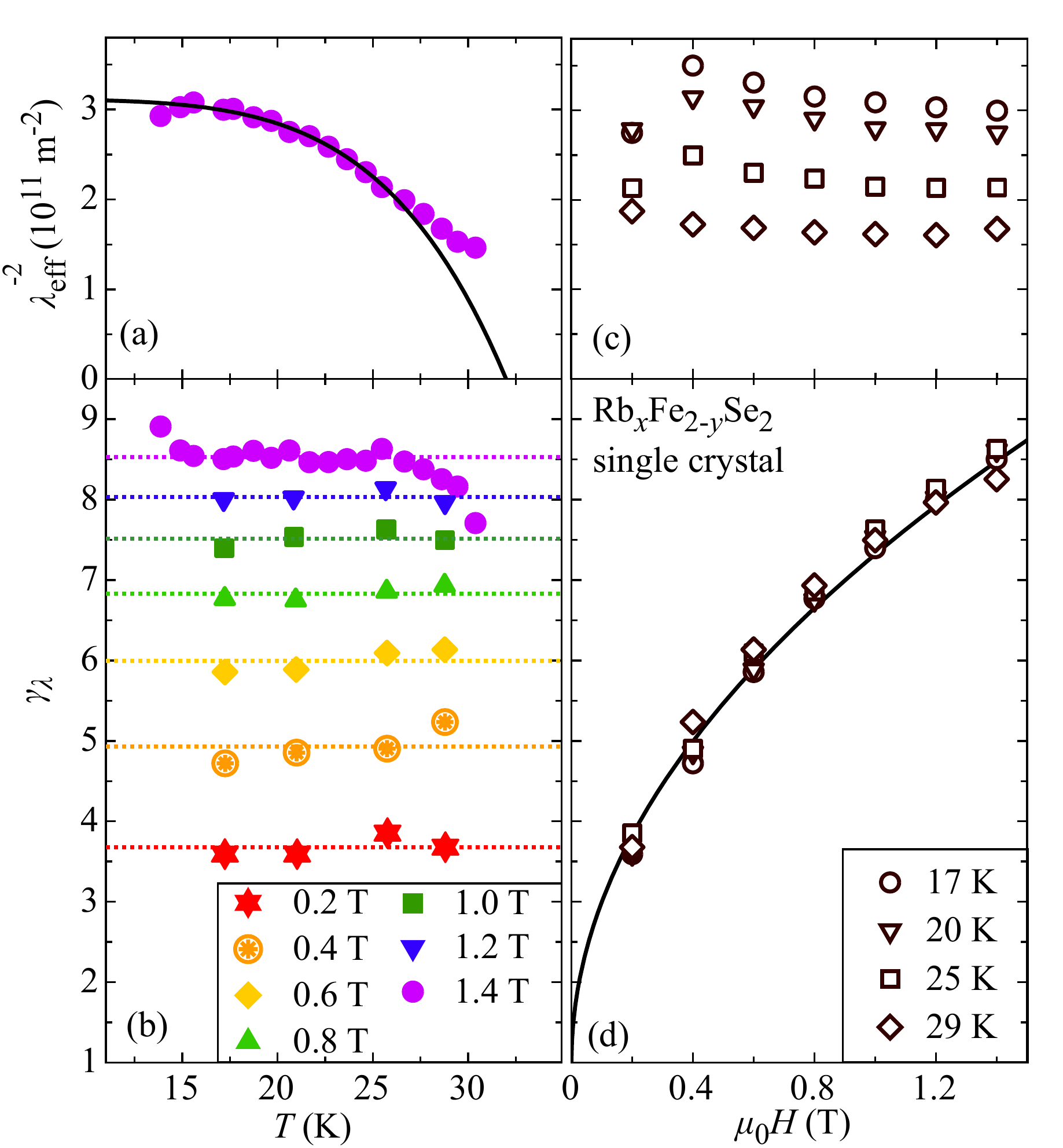}
\caption{(Color online) Summary of all the results obtained by analyzing the experimental torque signal of \RbFeSe\ with Eq.~(\ref{kogan}). (a) Temperature dependence of $\lambda_{\rm eff}^{-2}$. The line is a power law fit to the data at 1.4~T. (b) Temperature dependence of $\gamma_\lambda$ for various fields, showing that $\gamma_\lambda$ is strongly increasing with $H$, but is almost independent of $T$. The dotted lines represent the average $\gamma_\lambda$ for each field. (c) Field dependence of  $\lambda_{\rm eff}^{-2}$ for various temperatures. (d) Field dependence of $\gamma_\lambda$ for various $T$. The black line is a guide to the eye.}
\label{gamma_T}
\end{figure}      
\indent The temperature dependence of $\lambda_{\rm eff}$ presented in Fig.~\ref{gamma_T}a can be fitted with the empirical power law $\lambda_{\rm eff}^{-2}(T)=\lambda_{\rm eff}^{-2}(0)[1-(T/T_{\rm c})^n]$ with $T_{\rm c}\simeq32$~K, $n\simeq5.2$ and $\lambda_{\rm eff}(0)\simeq1.8$~$\mu$m. This rather large value of the effective penetration depth may be due to phase separation, as mentioned above. The value of $n$ is quite large compared to what is expected from the two-fluid model the fitting expression comes from. This might be related to phase separation, as the links between superconducting domains could depend on temperature, changing the temperature dependence of the penetration depth. Figure~\ref{gamma_T}b shows $\gamma_\lambda(T)$ for various $H$. It is almost temperature independent between $T_{\rm c}$ and $T_{\rm c}$/2 for all fields studied. The slight drop observed at higher temperatures in 1.4 T may be due to the proximity of the transition, which decreases the superconducting signal. The increase of $\gamma$ at 1.4 T at low temperatures is most likely due to pinning effects.\cite{Willemin1998a} Most importantly, a remarkable monotonous field dependence of $\gamma_\lambda$ is observed (Fig.~\ref{gamma_T}d), with the strongest dependence in the lowest fields. In low fields, $\gamma_\lambda({\rm 0.2~T})\simeq 3.5$. An extrapolation of the measured values of $\gamma_\lambda$ towards zero field yields $\gamma_\lambda(0~\rm T)\sim 1-2$. In Fig.~\ref{gamma_T}c the field dependence of $\lambda_{\rm eff}(H)$ is shown for various $T$. The effective penetration depth $\lambda_{\rm eff}(H)$ tends to a constant value for $\mu_0H \geq$ 1~T for all studied $T$.

\section{Discussion}

\indent The extrapolated $\lambda_{\rm eff}(0)\simeq 1.8$~$\mu$m is surprisingly large compared to that of other iron-based superconductors. In the related iron selenide FeSe$_{0.5}$Te$_{0.5}$ a much lower value of $\lambda_{\rm eff}(0)\simeq 0.7$~$\mu$m was reported.\cite{Bendele2010b} However, for \RbFeSe\ such a large $\lambda_{\rm eff}(0)$ is consistent with the very small $\mu_0H_{\rm c1}\lesssim 0.3$~mT observed in this work and in \KFeSe\ by others.\cite{Lei2011} The small $\gamma_\lambda(0~\rm T)\sim1-2$ at very low fields is also in agreement with an isotropic $H_{\rm c1}$. The increase of $\gamma_\lambda=\lambda_c/\lambda_{ab}$ with $H$ and the field independent $\lambda_{\rm eff}=(\lambda_{ab}^2\lambda_c)^{1/3}$ imply that $\lambda_{ab}=\lambda_{\rm eff}\gamma_\lambda^{-1/3}$ decreases with increasing $H$. This is consistent with the high-field NMR result.\cite{Torchetti2011} The small high-field value of $\lambda_{ab}$ reported in Ref.~\onlinecite{Torchetti2011}, in combination with the saturating $\lambda_{\rm eff}(H)$ and the field dependence of $\gamma_\lambda(H)$ observed here, suggests that $\gamma_\lambda$ continues to increase at higher fields. Note that the anisotropy of $\sim3$ (Ref.~\onlinecite{Mun2011}) observed in very high fields by upper critical field measurements cannot be directly compared with the magnetic penetration depth anisotropy $\gamma_\lambda$ investigated here, because in general $H_{\rm c2}^{||ab}/H_{\rm c2}^{||c}=\xi_{ab}/\xi_{c}=\gamma_\xi\neq\gamma_\lambda$.\cite{Weyeneth2009b}\\
\indent A field dependent $\gamma_\lambda$ might be associated with a complex band structure, since in the case of multiple superconducting gaps originating from different bands the superconducting screening currents, related to $\lambda_{ab}$ and $\lambda_{c}$, may give rise to an unusual behavior of $\gamma_\lambda$. A similar behavior was observed in MgB$_2$, where the two-gap excitation spectrum yields a strongly field dependent anisotropy.\cite{Angst2002, Angst2004} In MgB$_2$, as the field increases, the gap from the 3D $\sigma$-band is closed, so the main part of the superconducting fluid density comes from the 2D $\pi$-band. The 2D character of the remaining band implies a larger anisotropy. \\
\indent Band structure calculations\cite{Nekrasov2011, Shein2011, Cao2011} in \AFeSe\ yield multiple bands, with four 2D sheets on the sides of the Brillouin zone and one cylindrical sheet at the center. Depending on doping, this cylinder can be split into two 3D cones in K$_{0.8}$Fe$_2$Se$_2$,\cite{Nekrasov2011} which are completely detached (3D) in the calculation of Ref.~\onlinecite{Shein2011}, although when the authors use experimental lattice parameters these cones are replaced by a cylinder (2D). According to Ref.~\onlinecite{Cao2011}, the center band is a cylinder. In CsFe$_2$Se$_2$ and Cs$_{0.8}$Fe$_2$Se$_2$, this band consists in two almost detached cones.\cite{Nekrasov2011} In Tl$_{0.58}$Rb$_{0.42}$Fe$_{1.72}$Se$_2$,\cite{Mou2011} two gaps of different amplitudes were observed on the side and center bands, but the candidate for a 3D inner band was to small for a gap to be observed. It is therefore not yet clear whether for \RbFeSe\ with the doping used in this work a 3D band is present. Clearly, more material specific work is needed in order to clarify the interplay between multiband superconductivity and the anisotropy of \RbFeSe. However, if a 2D-3D band scenario is the origin of the field dependence of the anisotropy of \RbFeSe, it must be temperature dependent as well,\cite{Angst2002} which is not the case according to Fig.~\ref{gamma_T}d. \\ 
\indent The temperature independent $\gamma_\lambda$ suggests that the origin of its field dependence is not related to the superconducting gap energy, which is strongly temperature dependent in the examined temperature range. However, there is one energy scale which is almost temperature independent in the superconducting state: the N\'eel temperature $T_{\rm N}\approx 500$~K.\cite{Liu2011} A superconductor coexisting with an antiferromagnetic phase is expected to behave in a peculiar way, although the temperature range studied here is far too low to excite any change in this strongly coupled antiferromagnet. However, magnetic field modifications can lead to changes in the domain structure, and therefore changes of coupling between superconducting areas. This could result in variations of the “averaged” effective anisotropy. As was shown for various iron-based superconductors, the lattice parameters, in particular the pnictogen height in the iron-pnictides, are directly related to superconductivity.\cite{Mizuguchi2010} Importantly, such scaling also works for the iron-selenide layer.\cite{Krzton-Maziopa2011, Mizuguchi2010} It is also possible that magnetostrictive effects, which are expected to increase with magnetic field, may influence the lattice parameters and by that the anisotropy of the system. In such a case the strong curvature of $H_{\rm c2}$($T$) in the vicinity of $T_{\rm c}$ often observed in pnictides/chalcogenides may be explained by the change of $T_{\rm c}$ with $H$ caused by the change of pnictogen height.\\

\section{Summary}

\indent In summary, we present an investigation of the magnetic properties of \RbFeSe\, revealing a strong field dependence of $\gamma_\lambda$ ranging from $\gamma_\lambda(0.2~{\rm T})<4$ to $\gamma_\lambda(1.4~{\rm T})>8$. This behavior stands out among other iron-based superconductors, consistent with the singular coexistence of magnetic ordering and superconductivity. In accordance with lower critical field measurements, our data suggest that in very low fields $\gamma_\lambda(0~\rm T)\sim1-2$. At 1.4~T the effective magnetic penetration depth is $\lambda_{\rm eff}(0)\simeq1.8~\mu$m. The vortex phase in \RbFeSe\ is best described by an isotropic three dimensional state in low fields which becomes strongly anisotropic with increasing field. In this respect the novel iron-selenide \RbFeSe\ could be a potential candidate for magnetic-field tuned applications of superconductivity.\\

\section{Acknowledgements}

\indent This work was supported by the Swiss National Science Foundation, in part by the NCCR program MaNEP and the
Sciex-NMSch (project code 10.048), and by National Science Centre (Poland) based on decision No. DEC-2011/01/B/ST3/02374.


%

\end{document}